# Defects in graphite engineered by ion implantation for the self-assembly of gold nanoparticles

Yumeng Liu[1#], Yanhao Deng[1#], Yizhuo Wang[1], Li Wang[2], Tong Liu[2], Wei Wei[2], Zhongmiao Gong[2], Zhengfang Fan[1], Zhijuan Su[3*], Yanming Wang[1*], Yaping Dan[1*]

[1] University of Michigan – Shanghai Jiao Tong University Joint Institute, Shanghai Jiao Tong University, Shanghai, 20240 China

[2] Vacuum Interconnected Nanotech Workstation, Suzhou Institute of Nano-Tech and Nano-Bionics, The Chinese Academy of Sciences, Suzhou 215123, China[3] Global Institute of Future Technology, Shanghai Jiao Tong University, Shanghai, 20240 China

Email: zhijuan.su@sjtu.edu.cn, yanming.wang@sjtu.edu.cn, yaping.dan@sjtu.edu.cn

Abstract

Defect engineering in two-dimensional (2D) materials is essential for advancing applications such as gas sensing, single-atom catalysis, and guided nanoparticle self-assembly, enabling the creation of materials with tailored functionalities. This study investigates ion implantation effects on highly ordered pyrolytic graphite (HOPG) surfaces, using scanning tunneling microscopy (STM) and density functional theory (DFT) simulations to identify distinct defect structures. High-energy heavy ions cause inelastic scattering, increasing surface damage, while gold atoms deposited onto defect sites preferentially form atomic clusters. Through focused ion beam techniques, spatially distributed defects were engineered, guiding the self-assembly of nanoparticles. This research highlights the precision of ion irradiation for modifying HOPG surfaces, with significant implications for catalysis, nanotechnology, and the development of functional materials with controlled nanoscale properties.

Introduction

Two-dimensional (2D) materials, characterized by their atomically thin layers and the absence of dangling bonds on their surfaces, present exceptional opportunities for controlled defect engineering[1–5]. This distinctive feature ensures a high degree of surface stability while permitting precise defect engineering. The defect engineering on 2D materials is crucial for advancing applications in gas sensing[6–8], single-atom catalysis[9] and other applications[10–12]. Additionally, the ability to manipulate surface defects enables the guided self-assembly of nanoparticles or atomic clusters[13,14], paving the way for the development of advanced materials with precisely tailored functionalities. Gold nanoparticles (Au NPs) have emerged as the most extensively studied among noble metal nanoparticles[15–18]. Their excellent biocompatibility, diverse surface modification capabilities, and unique optical properties have rendered Au NPs indispensable in biomedical research[19]. These nanoparticles are at the forefront of innovations in medical imaging[20], tumor detection[21], and sensor development[22], highlighting the synergy between material science and biomedical applications.

In this context, highly ordered pyrolytic graphite (HOPG) stands out as a material of critical importance, particularly in surface catalysis[23], biomaterials[24], and fuel cell anodes[25]. However, to fully leverage HOPG's potential in these applications, a key challenge must be addressed: the controlled modification of its surface. Such modifications are essential for enhancing the HOPG functionality and ensuring its effective integration into various technological fields. Ion irradiation presents a powerful and precise technique for material modification[26]. Recent studies have demonstrated that low-energy heavy ion irradiation can effectively modify surfaces by precisely controlling the density and location of defects. For instance, 1-keV Ar+ ions were used to irradiate HOPG, resulting in surface defects concentrated within a few atomic layers[27]. Additionally, HOPG partially irradiated with 70 keV C+ ions at room temperature revealed nano-sized dendritic protrusions, with Raman spectroscopy indicating a phase transformation [28]. By carefully selecting the ion implantation energy and type of ions, researchers can fine-tune the properties of HOPG, introducing defects or creating nanostructures at atomic-level precision[27–30]. This method not only enhances the performance of these materials but also expands their application potential, driving innovation across multiple disciplines[31].

In this work, we systematically investigated the effects of ion implantation on the surface of a HOPG flake. By employing scanning tunneling microscopy (STM), we analyzed and identified three defect structures based on the distinct STM image characteristics and simulated STM images using density functional theory (DFT) calculations. The statistical distribution analysis shows that heavier ions implanted at a high energy are more likely to have inelastic scattering with the grahite lattice, resulting in more damages at the graphite surface. When gold is thermally evaporated onto these damaged surfaces, gold atoms aggregate at the defect sites, forming atomic clusters or nanoscale particles. This arregation is further supported by DFT calculations. We engineered a series of surface defects on graphite surface into a spatially distributed pattern by focused ion beam. These spatially distributed surface defects guide the self-assembly of atomic clusters or nanoscale particles into a similar pattern.

Results and Discussion

A fresh graphite surface was exfoliated from bulk highly ordered pyrolytic graphite (HOPG) by scotch tape. It was then implanted with boron ions at a fixed dose of $4 \times 10^{12}/cm^2$. The sample was imaged by scanning tunneling microscopy (STM) before and after ion implatation (as shown in Supplementary Information Fig.S1). Statistically, the defect density is approximately $2 \times 10^{12} /cm^2$ (refer to STM images of defects in Fig.S2), which is about half of the boron implantation dose of $4 \times 10^{12}/cm^2$. This indicates that a significant number of boron ions were implanted into the graphite surface without creating visible defects or creating defects deep inside that are invisible in STM. Excluding the possibility that multiple ions bombarded the surface at locations too close to be distinguishable by STM, it can be inferred that approximately half of the implanted ions did not create defects that are imageable by STM. Furthermore, a close examination reveals that the shapes and sizes of these defects vary considerably. It is unlikely that these differences are solely due to implantation by multiple ions, as some defects exhibit a broken lattice structure (Defect #1 in Fig.S1c), while others appear as bright bumps with an intact lattice, suggesting that the defects are located right below the surface (Defect #2 in Fig.S1c).

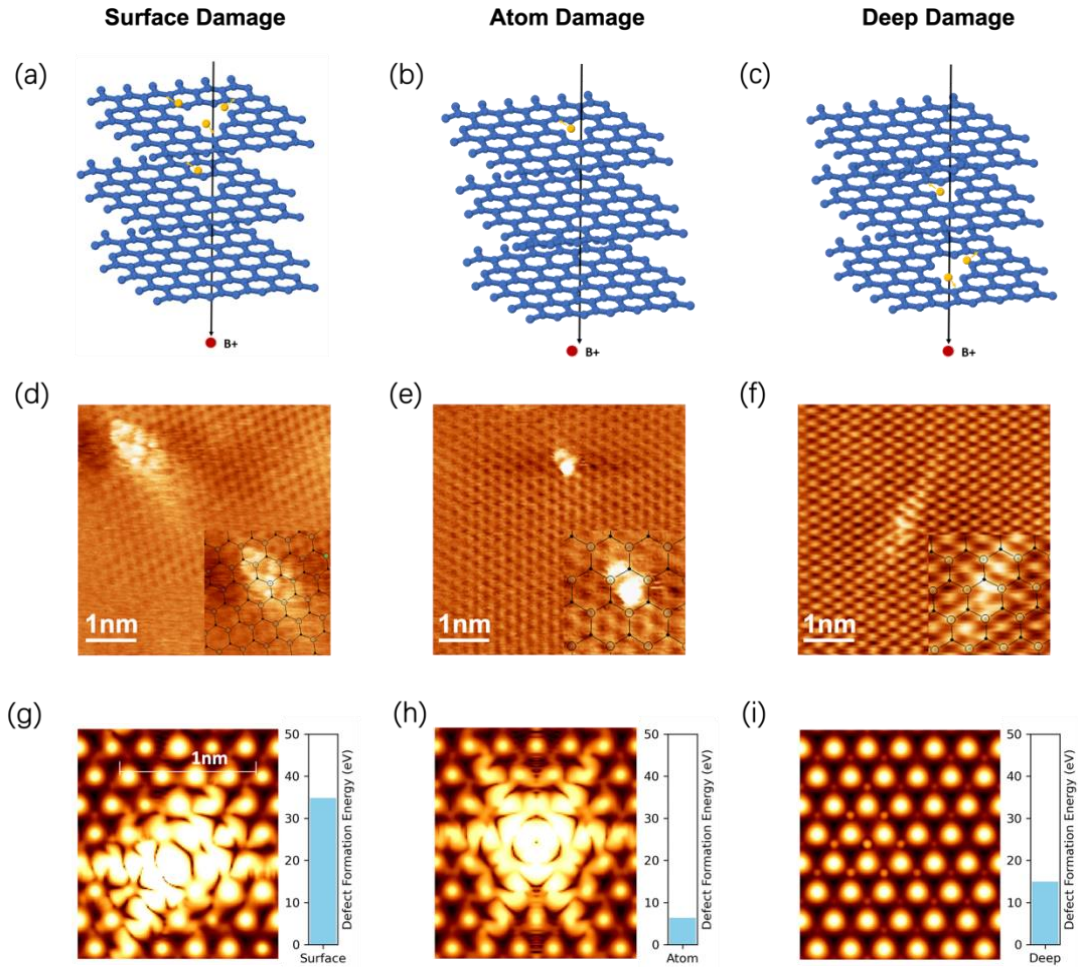

Figure.1 Three potential types of damage structures include (a)surface damage, (b)atom damage, (c) deep damage, their typical STM images in (d)-(f), scan parameters: (d) V = + 190.4 mV, I = 0.68 nA; (e) Vs =+ 80.6 mV, It = 1.39 nA; (f) Vs = + 153.5 mV, It = 1.03 nA. Their simulated STM images and defect formation energy calculated by DFT in (g)-(i).

Based on the STM images, we propose categorizing these defects into three distinct forms of lattice damage that single ions may induce on the surface of 2D materials. This classification is also supported by STM images simulated using first-principles calculations, a method that has been widely used and proven to be effective[32]. Fig.1a depicts the first type of lattice damage, characterized by the loss of multiple carbon atoms, potentially extending through several layers and forming a substantial cavity. This form of lattice damage may result from either large-scale damage caused by the direct impact of single ion implantation or from multiple ions implantation in close proximity. It manifests in STM images as a completely disordered arrangement of surface layer atoms, resulting in observable damages, as depicted in the typical STM image in Fig.1d. The coresponding lattice, automatically identified and generated by the WSxM software, reveals complete destruction, as illustrated in attached image. Fig.1g shows a simulated STM image of a graphite surface with multiple carbon atoms removed, obtained from DFT calculations. Similar to the experimental STM image, the simulation also exhibits a large bright bump with a disordered lattice. Fig.1b shows a defect characterized by the loss of a single carbon atom in the surface layer, while the underlying layers remain intact. This type of defects comes from localized damages caused by single ion-implantation or ion secondary sputtering. It appears as a single bright spot surrounded by an intact lattice structure, as shown in the typical STM image in Fig.1e. The associated lattice is aslo shown in the attached image, where the damaged area is confined within a single crystal cell. Consistently, the simulated STM image in Fig.1h shows that such a point defect exhibits a symetric atomic-scale pattern. Fig.1c illustrates defects indicative of deep damage, characterized by an intact lattice structure in the surface layer but with lattice damage occuring beneath. This type of defect arises when an ion penetrates the surface layer without causing visible damage but induces atom loss in the underlying layers. The experimental STM image (Fig.1f) shows a bright bump with a clear lattice structure on the graphite surface, a phenomenon commonly reported in the literature[27,33]. As shown in the attached image, in Fig.1e, the atoms remain in their crystal lattice arrangement, with only a slight elevation in height. The simulated STM image (Fig.1i) similarly shows a clear lattice but with a less pronounced bump in comparison with the experimental STM image. The formation energies for these three types of defects were also calculated with DFT, giving the values of 6.39 eV, 14.97 eV, and 34.85 eV, which are graphically shown on the right side of Fig. 1g, 1h, and 1i respectively. For these calculations, representative configurations were selected for each type of the defect (Fig. S2).

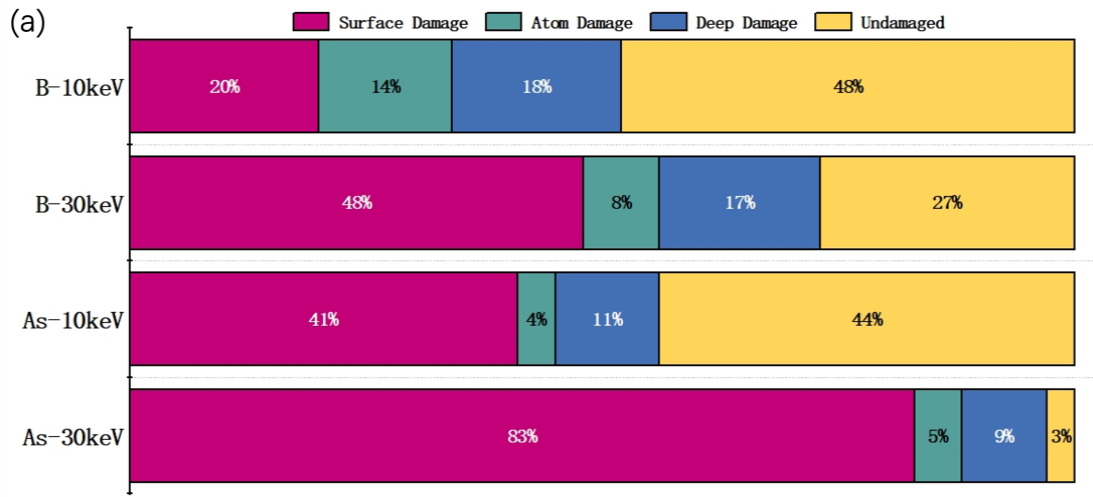

Figure.2 Statistics of the three types of damages and the proportion without defects.

Our experimental data show that the type of defects generated upon ion implantation into graphite is highly depedent on the mass and implanation energy of the incident ions. To investigate the probablity of these defects, we analyzed approximately one hundered STM images (refer to SI Fig. S3-S6 for details) and categorized them into three forms of lattice damage as depicted inFig.1. The statistical analysis is summarized in Fig.2a. We selected incident ions of varying masses, such as the commonly used $As^+$ and $B^+$ ions, with implantaion energies ranging from 10 keV to 50 keV, while maintaining a fixed dose of $4\times10^{12}$ cm$^{-2}$. Samples implanted at 50 kV were excluded from analysis due to the extensive surface damage, which made it challenging to distinguish between different forms of damage.

The distribution of the three defect types reveals that, except for the 10 keV $B^+$ ions, which produce a balanced proportion of the three defect types on the graphite surface, ion implantation under other conditions primarily results in surface damage. The proportion of surface damage increases with both the energy and diameter of the ions. Specifically, for the same type of ion, an increase in energy leads to an increase in surface damage and a decrease in the proportion of undamaged areas, while the proportions of atomic and deep damage decrease slightly. As we have calculated, the order of defect formation energy from small to large is atom damage, deep damage, and surface damage. Since the injection energy is significantly greater than the energy required to form the defect, an increase in injection energy makes surface damage more likely to occur. At constant energy, the size of the incident ions significantly affects defect formation. As$^+$ ions cause more substantial defects on the graphite surface compared to B$^+$ ions, both in quantity and size. This can be attributed to the lower momentum of B$^+$ ions with reduced energy and mass, resulting in less disturbance to the electron cloud and an increased likelihood of penetrating through surface atoms.

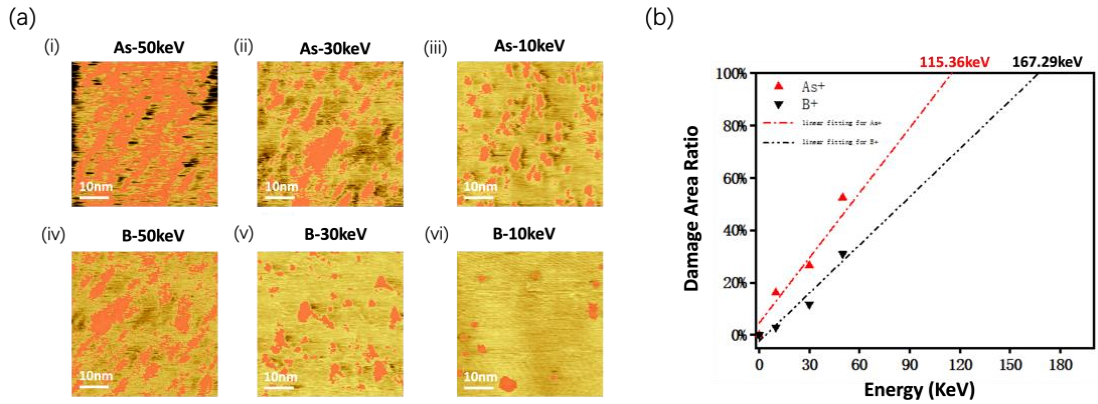

Figure.3 (a) Defect area on the HOPG irradiated by B$^+$ and As$^+$ vary from 10-50keV. (b) the damage area ratio of different energy.

Furthermore, we aim to explore the influence of defect size on the energy and type of the implantation source, which corresponds to the damage cross-section produced by a single ion in the lattice. During ion-solid interaction, the kinetic energy of incident ions is dissipated through elastic and inelastic collisions with the lattice. In elastic collisions, the lattice remains intact, and the energy is disspated as phonons or even photons. In inelastic collisions, the lattice is damaged, resulting in lattice deformations that can be imaged by STM. These deformations manifest as anomalous protrusions in STM images, a phenomenon that has been widely reported[27,34–38]. Therefore, we performed an area proportion analysis to derive the relationship between defect area ratio and energy, as illustrated in Fig.3. The results clearly demonstrate that as ion energy increases, the defect area on the HOPG surface proportionally increases. Linear fitting of the injection energy and defect area for B$^+$ and As$^+$ ions was performed, yielding predicted energies of 167.29keV and 115.36keV for 100% defect area, respectively.

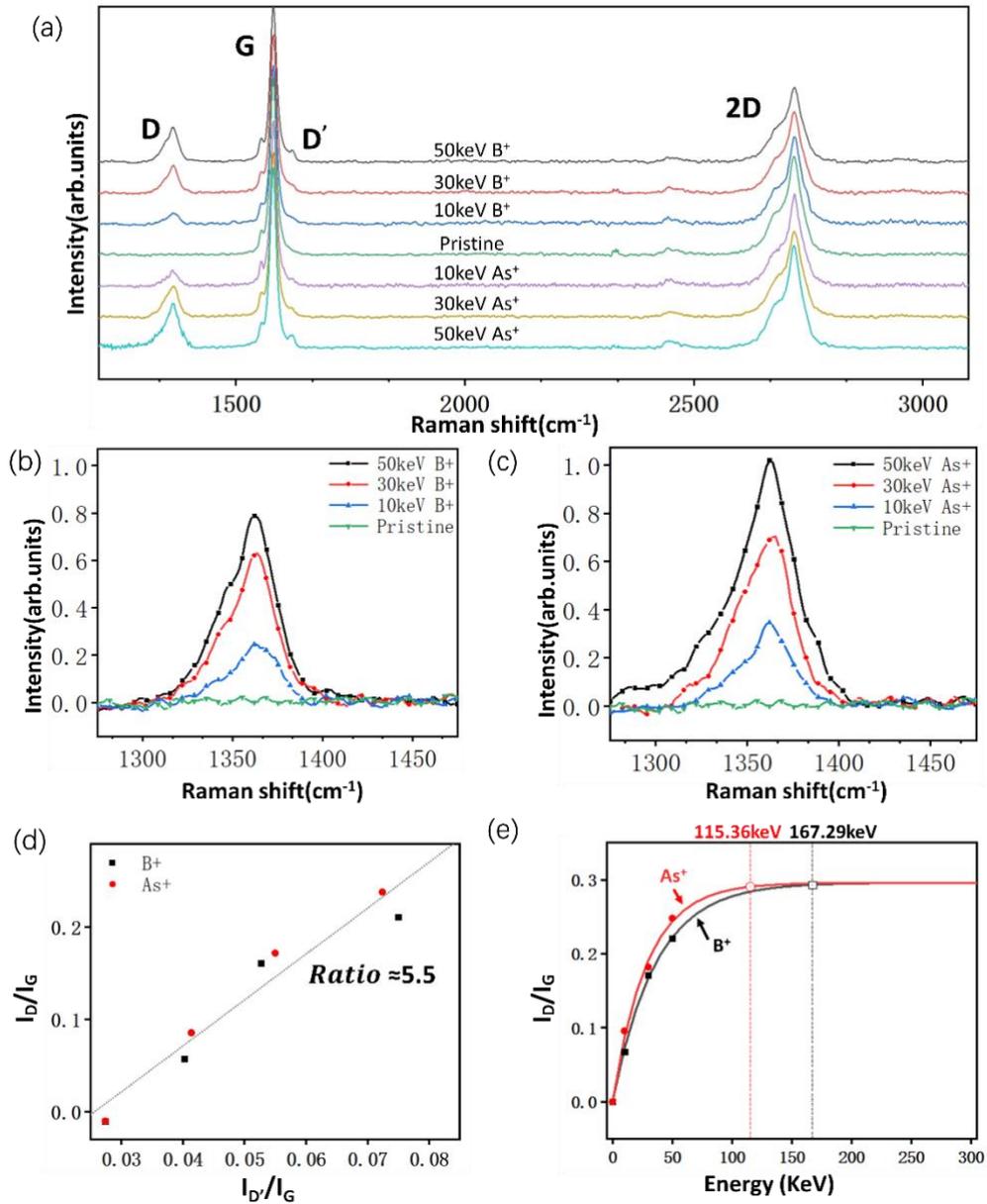

Figure.4 (a) Raman spectroscopy and the details of the D peak for (b) $B^+$ and (c) $As^+$; (d) $I_D/I_G$ versus ratio $I_{D'}/I_G$ and (e)The data is fitted according to the relation $f(\varphi)=\alpha[1-e^{-\sigma_0\varphi\cdot E}]$.

Raman spectroscopy has been widely used to identify damage in HOPG due to its sensitivity to changes in the vibrational modes of the material's lattice structure[39]. Ion implantation disrupts the periodic arrangement of carbon atoms, leading to defect formation, which alters the phonon modes and can be detected as changes in the Raman spectrum[40–44].

In pristine HOPG, the Raman spectrum is characterized by two prominent peaks: the G peak and the 2D peak, shown in Fig.4(a). The G peak, located around 1580 cm$^{-1}$, corresponds to the E$_{2g}$ phonon mode at the Brillouin zone center, associated with the in-plane vibrations of sp$^2$-bonded carbon atoms[41]. The 2D peak, around 2700 cm$^{-1}$, is a second-order overtone of the D peak and is indicative of the pristine nature of the graphite lattice. When the HOPG lattice is damaged, additional Raman features appear, most notably the D peak, located around 1360 cm$^{-1}$. The D peak arises from the breathing modes of sp$^2$ atoms in rings, requiring a defect for activation[45,46]. It results from a double resonance process involving phonons near the K point of the Brillouin zone. The intensity of the D peak increases with the level of disorder, making it a reliable indicator of lattice defects. Additionally, the D' peak, appearing around 1620 cm$^{-1}$, provides further insight for the nature of defects, specifically those related to edges and boundaries of damaged regions.

The ratio of the intensities of the D peak to the G peak ($I_D/I_G$) is commonly used to quantify the level of disorder in HOPG[43]. A higher $I_D/I_G$ ratio indicates a greater degree of disorder. Eckmann [47] noted that the $I_D/I_{D'}$ ratio can distinguish which defect types in graphitic materials it is: a maximum value of 13 for sp$^3$ hybridization defects, ~ 7 for vacancy-like defects, and a minimum of 3.5 for boundary defects. An $I_D/I_{D'}$ ratio of approximately 5.5 in Fig. 4(d) indicates that HOPG predominantly features vacancy-like and boundary defects. Tuinstra[44] reported a method to determine average crystal domain size by analyzing the $I_D/I_G$ intensity ratio, correlating this ratio with crystal domain dimensions to assess crystallinity and structural disorder in graphitic materials. Moreover, Mathew[40] proposed the fitting formula for the trend of $I_D/I_G$ with ion dose variation. Here, we point out that the ion energy also adheres to this formula, as shown in the following equation:

$$I_D/I_G = \alpha[1 - e^{-\sigma_0 \varphi \cdot E}] \qquad (1)$$

Where $\alpha$ is a fixed fitting parameter, $\sigma_0$ represent a species-specific parameters associated with the ions used for irradiation. E represents the energy of the ions and $\varphi$ is the radiation dose. Notably, $\sigma_0 \cdot E$ can be used to represent the single ion damage cross-section[40], the fitted parameters in Fig 4. (e) are shown in Table 1:

Table 1 Fitted Parameters in Fig 4. (e)

| Types of Ions | $\varphi$ | $\alpha$ | $\sigma_0$ |
|---|---|---|---|
| $As^+$ | $4 \times 10^{12}/cm^2$ | $0.29583 \pm 0.032$ | $8.677 \times 10^{-14} cm^2/keV$ |
| $B^+$ | $4 \times 10^{12}/cm^2$ | $0.29825 \pm 0.021$ | $6.954 \times 10^{-14} cm^2/keV$ |

Moreover, the $I_D/I_G$ ratio obtained from ion-impacted graphene saturates when the energy surpasses a specific value. Fig. 3(b) predicts that for a given dose, $B^+$ and $As^+$ ions will reach maximum surface defect proportions at 167.29 keV and 115.36 keV, respectively. These predicted energies are plotted in Fig. 4(e), where their voltages align with the onset of the saturation region, indicating that these levels correspond to the maximum cross-section damage. Beyond these points, the HOPG surface achieves maximum defect area, and higher energies do not further increase the $I_D/I_G$ ratio, explaining the saturation behavior of the curve regardless of dose or energy.

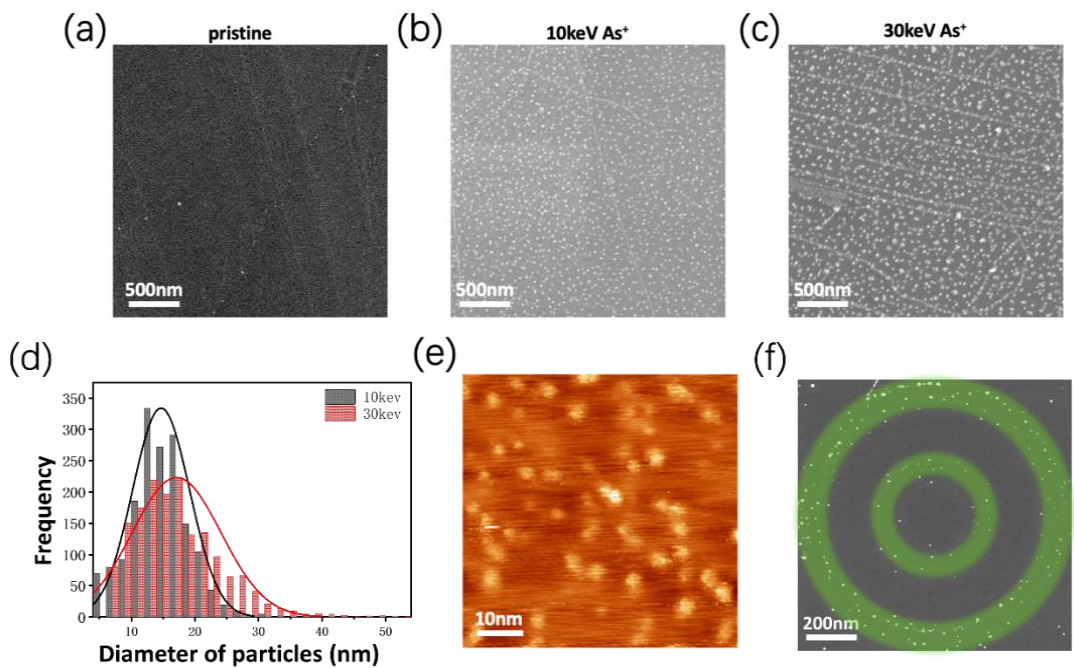

Figure.5 (a) SEM images for thermal evaporated 0.05 ML gold on pristine HOPG and HOPG implanted by (b) 10 keV and (c) 30 keV $As^+$ ions; (d) nanoparticle size distribution of gold on HOPG; (e) STM image of implanted by 10keV $As^+$ HOPG after thermally evaporated gold; (f)SEM images of gold particles arranged around regular damages.

The absence of dangling bonds on the surfaces of two-dimensional single-atom layer materials provides a unique advantage for the controlled defects we have engineered. These controlled defects can be pivotal for applications such as the guided self-assembly of nanostructures and single-atom catalysis. In this work, we aim to understand how these surface defects influence the formation of gold nanoparticles by evaporating 0.05 monolayers (ML) of gold onto both pristine and ion-implanted HOPG surfaces. It is illustrated in Fig. 5(a) that gold nanoparticles rarely form on the pristine HOPG surface, with only a few clusters appearing in the interlayer gaps. However, this observation changes dramatically when the HOPG surface is ion-implanted. The introduction of numerous defects on these surfaces, as shown in Figs. 5(b) and 5(c), significantly enhances the possibility to trap gold atoms, thereby facilitating the formation of gold nanoparticles. To quantify this effect, we conducted a detailed particle size analysis of the gold nanoparticles formed on both 10 keV and 30 keV $As^+$ ion-treated HOPG surfaces, as presented in Fig. 5(d). The analysis reveals that the gold nanoparticles on the HOPG surface treated with 10 keV $As^+$ ions are relatively uniform in size, with few large clusters. In contrast, the 30 keV $As^+$ ion-treated surface exhibits a broad size distribution of nanoparticles and a notable presence of large gold clusters exceeding 40 nm in diameter. This variability in nanoparticle size is directly related to the defect size distribution induced by ion implantation, highlighting the importance of ion energy and type in tailoring surface properties for specific applications. Fig. 5(e) shows an STM image of the HOPG surface implanted with 10keV $As^+$ ions after thermal evaporation of gold, which demonstrates the uniformity of the gold nanoparticles.

Furthermore, we created patterned defects on the HOPG surface to explore their potential for device fabrication. This experiment was conducted using a commercial focused ion beam (FIB). The $Ga^+$ ions used in FIB are very close in atomic number to the $As^+$ ions used in our experiments, making $Ga^+$ ions a suitable substitute for $As^+$ ions. We used FIB to fabricate two concentric rings on the HOPG surface. Fig. 5(f) shows the SEM image of gold particles aligned along the rings. The area of ion implantation is marked with green fluorescence in Fig. 5(f), representing two concentric circles, The clustering of gold particles along the rings is clearly visible, demonstrating the significant potential of this technique for orderly patterning and subsequent device fabrication processes.

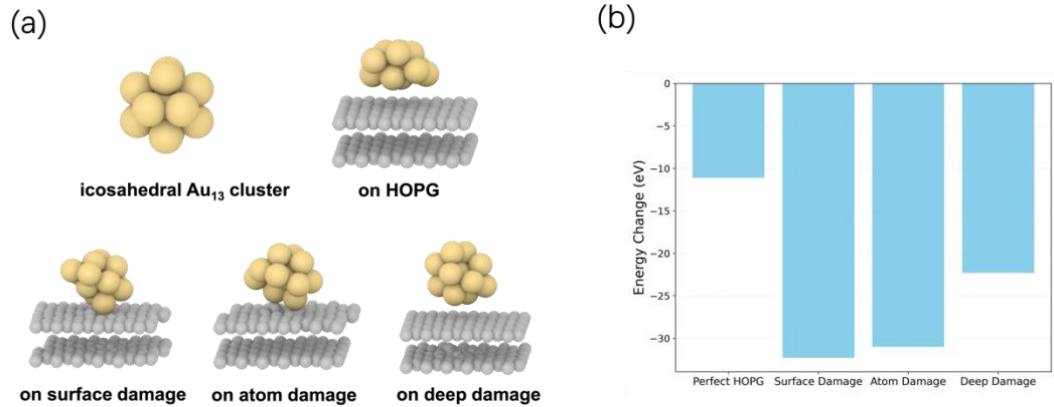

Figure 6: (a) Initial configuration of Au$_{13}$ cluster and the optimized structures of an Au$_{13}$ cluster placed on perfect HOPG, surface damage structure, atom damage structure, and deep damage structure. (b) The calculated adhesion energies of Au$_{13}$ cluster onto the graphite structures.

Finally, DFT calculations were conducted to provide atomistic insights into the device fabirciation process, predicting the contact geometries of Au cluster and HOPG substrate as well as the associated adhesion energies. Considering the high computational cost of DFT methods, the icosehedral Au$_{13}$ cluster was used as an example (as shown in Fig 6a), which is proposed to be the smallest stable 3-dimensional Au cluster [48]. Fig. 6a displays several DFT-optimized results that shows when an Au cluster lands on pristine and defective HOPGs, how it alters the local configurations, while Fig. 6b illustrates the corresponding energy changes. When attaching to a perfect graphite surface, the gold atoms tend to form a droplet shape. When the cluster approaches to surface surface defects, such as atomic or surface damage, it interacts with the dangling atoms, significantly lowering the energy of the configuration. However, when landing on the HOPG with deep damage, even though it may interact with the dangling atoms through an intermediate layer, the energy changes are less pronounced, only resulting in a slight reduction in the overall energy. Quantatively, these energy changes, which presumbaly can be understood as adhesion energies of Au$_{13}$ clusters, are -11.1eV, -32.3eV, -31.0eV, -22.3eV accordingly (Fig. 6b). This trend explains why higher injection energies, which create more surface damage, can enhance the trapping of gold clusters.

Conclusion

In summary, we have systematically explored the effects of ion implantation on the surface of a HOPG flake. Using scanning tunneling microscopy (STM) and density functional theory (DFT) simulations, we identified three distinct defect structures based on their unique STM image characteristics. Our statistical analysis reveals that heavier ions implanted at high energy are more prone to inelastic scattering with the graphite lattice, leading to increased surface damage. Additionally, when gold is thermally evaporated onto these damaged surfaces, gold atoms preferentially aggregate at the defect sites, forming atomic clusters or nanoscale particles. By employing a focused ion beam, we engineered spatially distributed surface defects, which successfully guided the self-assembly of atomic clusters or nanoscale particles into similar patterns. This work provides a deeper understanding of defect engineering on graphite surfaces and demonstrates the potential for precise patterning of nanostructures through ion implantation.

Experiment

Highly Oriented Pyrolytic Graphite (HOPG, 10 × 10 mm2, ZYA grade) was obtained from Xfnano Materials Tech (Nanjing, China), Several pieces were bought in order to repeat the experiments at least three times and check for reproducible results. Thus, three sets of seven samples were prepared, using six of them for irradiation and one of them as a reference (pristine). The fresh HOPG surface was exfoliated from a HOPG by scotch tape, which was then implanted. Ion irradiation was performed by 10-50 keV $B^+$ and $As^+$ with influence of $4 \times 10^{12}$ ions/cm$^2$ at room temperature (~ 25 °C) by the VIIsta810 ion implanter from Applied Materials at the Shanghai Institute of Technical Physics of the Chinese Academy of Sciences (SITP-CAS).

Then it was transferred to the STM chamber through a vacuum sealing device and baked at 200 degrees Celsius for 1 hour in ultrahigh vacuum (UHV) inside a preparation chamber with a base pressure of $2 \times 10^{-10}$ Torr to remove surface adsorbates. All experiments were conducted under high vacuum, and the samples were also kept in a vacuum state during transportation. The STM was performed in the Vacuum Interconnected Nanotech Workstation (Nano-X) at the Suzhou Institute of Nano-Tech and Nano-Bionics, Chinese Academy of Science. In situ scanning tunneling microscopy (STM) using a PtIr alloy tip, with 0.25 mm diameter and 8.0 mm length, was conducted to image the HOPG surface at the nano- or atomic scale.

The STM images of Defects in HOPG surface and analysis of gold particle size in SEM image were using Image J software. The Renishaw inVia confocal Raman microscope is applied to characterize the Raman and PL spectra of the MoS2, with a continuous wavelaser at 532 nm wavelength by the Instrumental Analytical Center (IAC), Shanghai Jiao Tong University. SEM and FIB are carried out by Thermo Scientific Scios 2 DualBeam, the minimum beam current of FIB is 1.6pA.

All DFT calculations were carried out using the Vienna ab initio simulation package (VASP) with the projected augmented wave (PAW) and Perdew-Burke-Ernzerhof (PBE) functional, employing a cut-off energy of 400eV. Van der Waals (vdW) interactions were included by the DFT-D3 method with Becke-Johnson damping to account for the interlay interactions [49]. The brillioin zone was sampled using a Γ-centered 2 × 2 × 1 k-point mesh. A vacuum layer of at least 13 Å was introduced along the z-direction to avoid self-interactions. The energy convergence criterion for the DFT self-consistent calculations was set to $10^{-6}$ eV. STM images were calculated using the $E_F \pm 0.5$ e, a method based on a simple model of an s-wave STM tip [50]. A bias voltage of $E_F \pm 0.5$ e was chosen for efficiencey, as other values were found to produce qualitatively similar images.

Acknowledgement

This work was financially supported by Oceanic Interdisciplinary Program of Shanghai Jiao Tong University (SL2022ZD107), Shanghai Pujiang Program (22PJ1408200), National Science Foundation of China (NSFC, No. 92065103), and Shanghai Jiao Tong University Scientific and Technological Innovation Funds (2020QY05). Ion implantation was performed at the Shanghai Institute of Technical Physics, Chinese Academy of Science. Focused Ion Beam was conducted at the center for NanoX at the Suzhou Institute of Nano-Tech and Nano-Bionics, Chinese Academy of Science. Raman spectrum and photoluminescence measurements were carried out at the Instrumental Analytical Center (IAC), Shanghai Jiao Tong University.

References

[1] M. Schleberger, J. Kotakoski, *Materials* **2018**, *11*, 1885.

[2] J. Jiang, T. Xu, J. Lu, L. Sun, Z. Ni, *Research* **2019**, *2019*.

[3] M. F. Hossen, S. Shendokar, S. Aravamudhan, *Nanomaterials* **2024**, *14*, 410.

[4] M. Telkhozhayeva, O. Girshevitz, *Adv. Funct. Mater. n/a*, 2404615.

[5] Z. Huang, H. Liu, R. Hu, H. Qiao, H. Wang, Y. Liu, X. Qi, H. Zhang, *Nano Today* **2020**, *35*, 100906.

[6] M. Sai Bhargava Reddy, S. Aich, *Coord. Chem. Rev.* **2024**, *500*, 215542.

[7] X. Fu, Z. Qiao, H. Zhou, D. Xie, *Chemosensors* **2024**, *12*, 85.

[8] M. Mathew, P. V. Shinde, R. Samal, C. S. Rout, *J. Mater. Sci.* **2021**, *56*, 9575.

[9] X. Wang, Y. Zhang, J. Wu, Z. Zhang, Q. Liao, Z. Kang, Y. Zhang, *Chem. Rev.* **2022**, *122*, 1273.

[10] C. Ataca, S. Ciraci, *J. Phys. Chem. C* **2011**, *115*, 13303.

[11] S. McDonnell, R. Addou, C. Buie, R. M. Wallace, C. L. Hinkle, *ACS Nano* **2014**, *8*, 2880.

[12] D. Ni, J. Zhang, W. Bu, H. Xing, F. Han, Q. Xiao, Z. Yao, F. Chen, Q. He, J. Liu, S. Zhang, W. Fan, L. Zhou, W. Peng, J. Shi, *ACS Nano* **2014**, *8*, 1231.

[13] XPS/STM study of model bimetallic Pd–Au/HOPG catalysts, *Appl. Surf. Sci.* **2016**, *367*, 214.

[14] X. Zhou, Q. Shen, K. Yuan, W. Yang, Q. Chen, Z. Geng, J. Zhang, X. Shao, W. Chen, G. Xu, X. Yang, K. Wu, *J. Am. Chem. Soc.* **2018**, *140*, 554.

[15] Y. Wu, L. Wang, M. Yan, X. Wang, X. Liao, C. Zhong, D. Ke, Y. Lu, *Adv. Healthc. Mater.* **2024**, *n/a*, 2400836.

[16] M. Zhang, S. Shao, H. Yue, X. Wang, W. Zhang, F. Chen, L. Zheng, J. Xing, Y. Qin, *Int. J. Nanomedicine* **2021**, *16*, 6067.

[17] Z. Deng, C. Yang, T. Xiang, C. Dou, D. Sun, Q. Dai, Z. Ling, J. Xu, F. Luo, Y. Chen, *J. Nanobiotechnology* **2024**, *22*, 157.

[18] N. Dridi, Z. Jin, W. Perng, H. Mattoussi, *ACS Nano* **2024**, *18*, 8649.

[19] S. Ahmad, S. Ahmad, S. Ali, M. Esa, A. Khan, H. Yan, *Int. J. Nanomedicine* **2024**, *19*, 3187.

[20] F.-Y. Kong, J.-W. Zhang, R.-F. Li, Z.-X. Wang, W.-J. Wang, W. Wang, *Molecules* **2017**, *22*, 1445.

[21] L. Li, Y. Wei, S. Zhang, X. Chen, T. Shao, D. Feng, *J. Electroanal. Chem.* **2021**, *880*, 114882.

[22] S. Mehmood, R. Ciancio, E. Carlino, A. S. Bhatti, *Int. J. Nanomedicine* **2018**, *13*, 2093.

[23] A. V. Bukhtiyarov, M. A. Panafidin, I. A. Chetyrin, I. P. Prosvirin, I. S. Mashkovsky, N. S. Smirnova, P. V. Markov, Y. V. Zubavichus, A. Yu. Stakheev, V. I. Bukhtiyarov, *Appl. Surf. Sci.* **2020**, *525*, 146493.

[24] Y. Gao, J. Chen, J. Liu, M. Li, Y. Wang, *Adv. Mater. Interfaces* **2024**, *n/a*, 2400557.

[25] Y. Zhou, T. Holme, J. Berry, T. R. Ohno, D. Ginley, R. O'Hayre, *J. Phys. Chem. C* **2010**, *114*, 506.

[26] Y. Zhang, W. J. Weber, *Appl. Phys. Rev.* **2020**, *7*, 041307.

[27] X. Wang, G. Li, L. Zhang, F. Xiong, Y. Guo, G. Zhong, J. Wang, P. Liu, Y. Shi, Y. Guo, L. Chen, X. Chen, *Appl. Surf. Sci.* **2022**, *585*, 152680.

[28] Y. Zhou, Y. Wang, Q. Lei, Q. Huang, W. Zhang, L. Yan, *Carbon Lett.* **2021**, *31*, 593.

[29] W. Ni, Q. Yang, H. Fan, L. Liu, T. Berthold, G. Benstetter, D. Liu, *J. Nucl. Mater.* **2015**, *464*, 216.

[30] Y. Jin, J.-M. Song, K.-B. Roh, N.-K. Kim, H.-J. Roh, Y. Jang, S. Ryu, B. Bae, G.-H. Kim, *J. Korean Phys. Soc.* **2016**, *69*, 518.

[31] B. Maharana, M. K. Rajbhar, G. Sanyal, B. Chakraborty, R. Jha, S. Chatterjee, *Electrochimica Acta* **2023**, *464*, 142868.
[32] J. M. Blanco, C. González, P. Jelínek, J. Ortega, F. Flores, R. Pérez, *Phys. Rev. B* **2004**, *70*, 085405.
[33] B. An, S. Fukuyama, K. Yokogawa, M. Yoshimura, *Jpn. J. Appl. Phys.* **2000**, *39*, 3732.
[34] E. Bourelle, Y. Tanabe, E. Yasuda, S. Kimura, *Carbon* **2001**, *39*, 1557.
[35] R. Coratger, A. Claverie, F. Ajustron, J. Beauvillain, *Surf. Sci.* **1990**, *227*, 7.
[36] I. C. Gebeshuber, S. Cernusca, F. Aumayr, H. P. Winter, *Int. J. Mass Spectrom.* **2003**, *229*, 27.
[37] J. C. M. López, M. C. G. Passeggi, J. Ferrón, *Surf. Sci.* **2008**, *602*, 671.
[38] K. Havancsák, L. P. Biró, J. Gyulai, A. Ju. Didyk, *Radiat. Meas.* **1997**, *28*, 65.
[39] X. Cong, M. Lin, P.-H. Tan, *J. Semicond.* **2019**, *40*, 091001.
[40] S. Mathew, T. K. Chan, D. Zhan, K. Gopinadhan, A.-R. Barman, M. B. H. Breese, S. Dhar, Z. X. Shen, T. Venkatesan, J. T. L. Thong, *Carbon* **2011**, *49*, 1720.
[41] J. Zeng, J. Liu, H. J. Yao, P. F. Zhai, S. X. Zhang, H. Guo, P. P. Hu, J. L. Duan, D. Mo, M. D. Hou, Y. M. Sun, *Carbon* **2016**, *100*, 16.
[42] Ion-induced damage in graphite: A Raman study, *J. Nucl. Mater.* **2010**, *403*, 108.
[43] L. Venosta, N. Bajales, S. Suárez, P. G. Bercoff, *Beilstein J. Nanotechnol.* **2018**, *9*, 2708.
[44] F. Tuinstra, J. L. Koenig, *J. Chem. Phys.* **1970**, *53*, 1126.
[45] H. Peng, M. Sun, D. Zhang, D. Yang, H. Chen, R. Cheng, J. Zhang, Y. Wang, W. Yuan, T. Wang, Y. Zhao, *Surf. Coat. Technol.* **2016**, *306*, 171.
[46] E. A. Franceschini, G. I. Lacconi, *Appl. Surf. Sci.* **2017**, *400*, 254.
[47] A. Eckmann, A. Felten, A. Mishchenko, L. Britnell, R. Krupke, K. S. Novoselov, C. Casiraghi, *Nano Lett.* **2012**, *12*, 3925.
[48] R. Arratia-Perez, A. F. Ramos, G. L. Malli, *Phys. Rev. B* **1989**, *39*, 3005.
[49] S. Grimme, S. Ehrlich, L. Goerigk, *J. Comput. Chem.* **2011**, *32*, 1456.
[50] V. Wang, N. Xu, J.-C. Liu, G. Tang, W.-T. Geng, *Comput. Phys. Commun.* **2021**, *267*, 108033.